# Investigating Degradation Modes in Zn-AgO Aqueous Batteries with *In-Situ* X-ray Micro Computed Tomography

Jonathan Scharf[1], Lu Yin[1], Christopher Redquest[2], Ruixiao Liu[1], Xueying L. Quinn[3], Jeff Ortega[4], Xia Wei[4], Joseph Wang[1,5], Jean-Marie Doux[1]*, Ying Shirley Meng[1,5]*

[1] Department of Nano-Engineering, University of California San Diego, La Jolla, California 92093, United States of America
[2] Department of Chemical Engineering, University of California San Diego, La Jolla, California 92093, United States of America
[3] Department of Materials Science and Engineering, University of California San Diego, La Jolla, California 92093, United States of America
[4] ZPower LLC, Camarillo, CA 93012, USA
[5] Sustainable Power & Energy Center (SPEC), University of California San Diego, La Jolla, California 92093, United States of America
* Corresponding authors: jdoux@eng.ucsd.edu (J.M.D.), shmeng@ucsd.edu (Y. S. M.)

**Keywords: Zinc, Zn-based batteries, X-ray computed tomography, *In-situ* characterization, Corrosion**

## Abstract

To meet the growing global energy demands, the degradation mechanisms of energy storage devices must be better understood to improve performances. As a non-destructive tool, X-ray Computed Tomography (CT) has been increasingly used by the battery community to perform *in-situ* experiments that can investigate dynamic phenomena. However, few have used X-ray CT to study representative battery systems over long cycle lifetimes (>100 cycles). Here, we report the *in-situ* CT study of Zn-Ag batteries and demonstrate the effects of current collector parasitic gassing over long-term storage and cycling. We design performance representative *in-situ* CT cells that can achieve >250 cycles at a high areal capacity of 12.5 mAh/cm$^2$. Combined with electrochemical experiments, the effects of current collector parasitic gassing are revealed with micro-scale CT (MicroCT). The volume expansion and evolution of ZnO and Zn depletion is quantified with cycling and elevated temperature testing. The experimental insights are then utilized to develop larger form-factor (4 cm$^2$) cells with electrochemically compatible current collectors. With this, we demonstrate over 325 cycles at a high capacity of 12.5 mAh/cm$^2$ for a 4 cm$^2$ form-factor. The results of this work demonstrate that *in-situ* X-ray CT used in long cycle-lifetime studies can be applied to examine a multitude of other battery chemistries to improve their performances.

# 1. Introduction

To address the growing energy concerns, high-performance and cost-effective energy storage solutions, such as batteries, need to be developed to enable technologies such as in grid storage, Internet of Things (IoT), and high power electronics.[1–3] While Li-Ion Batteries (LIBs) have received considerable attention due to their high energy densities, there are significant concerns surrounding the safety, recyclability, $CO_2$ footprint, and mining ethics of precious metals that warrant the development of alternative chemistries.[4–9] With the benefits of lower material costs, benign chemistry, nonflammability, ambient processability, and high theoretical energy density (820 mAh/$g_{Zn}$, 5,854 Ah/L), aqueous zinc batteries (aqZB)[10] are a promising alternative to LIBs.[11–15] Additionally, given the higher ionic conductivities of aqueous electrolytes (~1 mS/cm for non-aqueous LIB, ~100 mS/cm for aqZBs),[12,13] aqZBs demonstrate superior performances in high current and pulse discharge applications in areas such as electric vehicles[16] and high power electronics.[17,18] Moreover, the benign chemistry and ambient processability allow for the development of flexible and stretchable batteries with easily adjustable form factors for applications in health monitoring and wearable electronics.[17,19–21]

However, aqZBs suffer from poor cyclability and shelf-life, which originate from issues such as Zn anode shape change, dendrite growth, Zn dissolution, and parasitic gassing.[20,22–25] To improve performances, several strategies have been explored to stabilize the anode and limit dendrite growth and anode shape change, such as the incorporation of additives in the electrode and electrolyte *(e.g,* $Bi_2O_3$[26,27], CTAB[28], and polyethylene glycol (PEG)[13,23,29]) and the development of novel 3D Zn sponge structures[16,30,31]. Nevertheless, a deeper fundamental understanding of the anode degradation is needed to better improve performances.

The instability and corrosion of the Zn anode in basic electrolytes result in gassing and hydrogen evolution reactions (HER) (equations 1-2),[32,33]

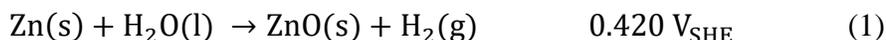
$$Zn(s) + H_2O(l) \rightarrow ZnO(s) + H_2(g) \qquad 0.420\ V_{SHE} \qquad (1)$$

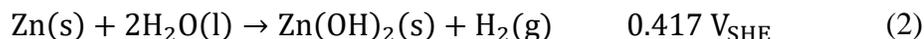
$$Zn(s) + 2H_2O(l) \rightarrow Zn(OH)_2(s) + H_2(g) \qquad 0.417\ V_{SHE} \qquad (2)$$

where $V_{SHE}$ refers to the voltage versus a standard hydrogen electrode. As a result, electrolyte additives such as ZnO and LiOH are often employed to limit hydrogen evolution reactions (HER) and anode corrosion.[13] Cathode materials in aqZBs tend to suffer from oxygen evolution reactions (OER) caused by high cathode potentials.[34] For instance, in Zn-Ag batteries, AgO offers a higher open circuit voltage (OCP) than $Ag_2O$ (1.86 V for AgO vs. 1.56 V for $Ag_2O$) at nearly double the theoretical capacity (430 mAh/g for AgO vs. 230 mAh/g for $Ag_2O$), and has demonstrated a high areal capacity of 54 mAh/cm² in a printable and flexible architecture.[17,35–37] However, as shown in equations (3-4), OER can facilitate AgO decomposition, resulting in capacity fade and self-discharge in Zn-AgO batteries.[38,39]

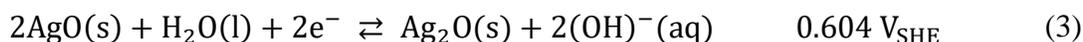
$$2AgO(s) + H_2O(l) + 2e^- \rightleftarrows Ag_2O(s) + 2(OH)^-(aq) \qquad 0.604\ V_{SHE} \qquad (3)$$

$$2(OH)^-(aq) \rightleftarrows H_2O(l) + \frac{1}{2}O_2(g) + 2e^- \qquad 0.401\ V_{SHE} \qquad (4)$$

$$2AgO(s) \rightarrow Ag_2O(s) + \frac{1}{2}O_2(g) \qquad (5)$$

OER can also increase the amount of dissolved oxygen in the electrolyte, facilitating the corrosion of the Zn anode.[40] Furthermore, both OER and HER of non-active materials in contact with the electrolyte, such as the current collectors, can limit the electrochemical performances of aqZBs[41]. As Bonnick and Dahn showed, the parasitic gassing of the current collectors and casing in Ni-Zn coin cells can drastically influence the cycle life.[42]

To investigate these degradation mechanisms, X-ray CT has been increasingly used in zinc-based battery research as a tool to probe and quantify valuable 3D morphological parameters, such as porosity and particle distribution.[43–45]. For example, CT has been used to show how the solid volume fraction in 3D porous Zn structures can improve the Zn Depth-of-Discharge ($DoD_{Zn}$) and energy density of anodes.[43]. Additionally, CT is ideal for *in-situ* or *operando* studies due to its non-destructive nature, and since grey values in images are proportional to the material's X-ray attenuation, species can often be separated to observe the chemical evolution in electrodes. Franke-Lang *et al.*, for example, were able to distinguish Zn from ZnO, and used *in-situ* CT to visually investigate the degradation and volume expansion of a zinc-air battery.[43] Likewise, Tobias *et al.* used o*perando* CT to reveal Zn particle depletion to follow a core-shell model, in which ZnO continuously grows around and depletes the Zn particle core.[44]

Most *in-situ* X-ray CT studies of battery systems are limited to only the first few cycles,[46,47] with few reports of long cycle life *in situ* CT studies (>100 cycles) that are representative of larger format cells and practical battery operation.[48] Herein, we used *in-situ* micro-scale X-ray CT (MicroCT) to study the effects of parasitic gassing on the performance of printed Zn-AgO batteries. As detailed in **Figure 1**, we developed *in-situ* CT cells that are representative of previously reported cell performances[17] to provide high resolution scans and detailed analysis of the battery system. This CT cell design allows for longer cycling (> 250 cycles) and was used to investigate the Zn particle development and growth of ZnO *in-situ* with cycling. Finally, we showed how the insights of this study can be applied to larger scale devices by demonstrating improved shelf and cycle life of a printed Zn-AgO battery.

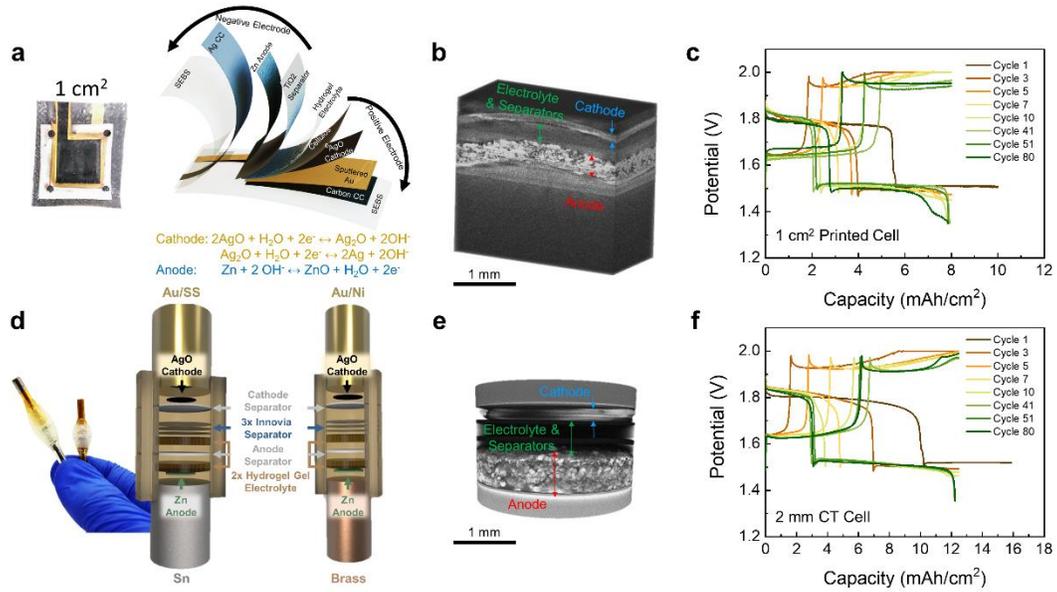

**Figure 1 | Comparison of the cell design and performances of the Zn-AgO printed cell (top - adapted from Ref [17]) and *in-situ* CT cell (bottom).** Comparison of cell architecture (a, d), 3D X-ray CT tomographic reconstructions (b, e), and electrochemical performances (c, f) for Zn-AgO printed cell (top row) and *in-situ* CT cell (bottom row).

## 2. Experimental Section

*Fabrication and Synthesis of Cell Components*

The electrodes were fabricated in similar ways compared with those in our previous work.[17] A binder resin was prepared by dissolving 1 g GBR-6005 fluorine rubber in 2.75 g acetone. The cathode slurry was prepared by mixing 0.95 g AgO powder and 0.05 g Super-P powder, followed by adding 1.25 g binder resin and mixing in a planetary mixer (Flaktak SpeedmixerTM DAC 150.1 FV) at 2500 rotations per minute (rpm) for 10 min. The anode slurry was prepared by mixing 0.91 g Zn powder and 0.09 g $Bi_2O_3$ powder, followed by adding 0.25 g binder resin and mixing in a planetary mixer at 1800 rpm for 5 min. Both slurries were then casted using a doctor blade, cured at 80 °C for 30 min, and trimmed into the desired shape for later use.

The synthesis of the PVA hydrogel is adapted from our previous work.[17] A PVA solution was prepared by dissolving 1 g of PVA in 5 g of deionized water, and a hydroxide solution was prepared by dissolving 0.2g $Ca(OH)_2$, 0.109 g LiOH and 5 g KOH in 26.159 g deionized water and removing any precipitated solid. The hydroxide solution was then added to the PVA solution in a 15.775: 10 weight ratio while stirring to obtain the hydrogel precursor. The precursor was then placed in a Patri dish in a vacuum desiccator to dry and crosslink until only 27.62% of the original weight was left. The gel was then cut to the desired size and stored in a 10 M KOH + 0.6 M LiOH solution at 20 °C.

*CT Cell Construction*

The *in-situ* cells were constructed using PTFE tubing with a 4 mm and 3 mm inner diameter for the Sn-Au/SS and Brass-Au/Ni respectively. The following metallic current collector rods were purchased from McMaster-Carr: 18-8 stainless steel, grade 2 titanium, 400 nickel, 360 brass, and 110 copper. The Sn rods were fabricated in-house by melting lead free solder (99% Sn, .3% Ag, .7% Cu) and using the PTFE tubes as molds for making 4 mm and 3 mm inner diameter rods. Gold electroplating was performed on the cathode rods using commercial plating solutions.[49] Before plating, the rods were first electro-cleaned at -2 V vs. stainless steel for 2 minutes. The stainless-steel rods then underwent an extra nickel strike step for 2 minutes at -2 V vs. stainless steel to facilitate the adhesion of the Au. Lastly, the rods were electroplated using a potassium aurocyanide solution by applying -1.5 V vs. a platinum counter electrode for 20 minutes. The calculated thickness from the coulomb count and the surface area of the rods was determined to be ~500 nm. A picture of all the current collector rods is presented in **Figure S1a**.

The active material was punched smaller than the current collectors, with a 2 mm diameter, and placed directly onto the rods. For a direct comparison of the device performance, both devices in the *in-situ* study of **Figure 3** used the same batches of anode and cathode material. Similarly, the active materials used in the CT cells of **Figure 5** were of the same batch. Commercial separators were supplied by ZPower LLC.: (1) Flexible Alkaline Separator (FAS) film that was used as the anode separator, consisting of a microporous polyolefin film with an inorganic filler[50], (2) a cellophane-based film (Innovia Films Company) that was used to help block silver migration,[51,52] and (3) a porous microporous EVOH film to separate the cellophane separator from the cathode. The order of layers for the *in-situ* cells is outlined in **Figure 1d**. The hydrogel electrolyte and separators were punched to the size of the tube diameter, larger than the active material to prevent crossover of Zn and Ag species around the edges. A 60% weight ratio of gel electrolyte to active material was found to properly wet electrodes for adequate cycling. The anode-cathode ratio was kept at approximately 5:1 to cycle the anode at ~10% $DoD_{Zn}$. For consistency, the same weight percentages and ratios of the active material and electrolyte were used for the $4cm^2$ form factor cells. After assembly, commercial epoxy (Gorilla Glue) was used to seal the cells from end to end to avoid electrolyte leakage.

*$4cm^2$ Form-Factor Cell Construction*

The larger form factor cells were constructed with 4 $cm^2$ metal foil current collectors with a 0.4 cm × 3 cm tab to allow for electrical contact (See **Figure S5a**). The foils were placed on a Styrene-ethylene-butylene-styrene (SEBS) thermo-elastomer substrate with epoxy (Gorilla glue) applied to the tab regions to allow for better sealing of the metals. Following this, the active materials were added atop the foils with 25 µL of liquid electrolyte (10M KOH, 0.6M LiOH) applied to the cathode to enhance wetting. Next, the PVA hydrogel electrolyte and separators were added. The anode and cathode stacks were then joined, and the SEBS sheets were vacuum- and heat-sealed (**Figure S5b**). A final layer of wear-resistant nylon (McMaster-Carr) was then added, and vacuum and heat sealed outside the SEBS sheets to ensure robust sealing. The final cell measured 3.5 x 3.5 $cm^2$ with a thickness of 2.05 mm, and an active area of 2 x 2 **$cm^2$**.

The following metal foils were used in the assembly of the 4 $cm^2$ cells and in electrochemical testing: 18-8 stainless steel (McMaster-Carr), 99.99% titanium (Amazon), nickel 200/201 (McMaster-Carr), 260 brass (McMaster-Carr), 99.99% Copper (MTI), and 99.99% Tin (Amazon) with thicknesses of 25-30 µm. For the gold coated foils, ~500 nm of Au was sputtered with a thin Ti adhesion layer at an Ar gas flow rate of 16 SCCM. A DC power of 200 W and 300 W was used for the Au and Ti sputtering respectively. The sputtering was performed with a Denton Discovery 635 Sputter System (Denton Discovery 635 Sputter System, Denton Vacuum, LLC, Moorestown, NJ, USA).

*Cell Cycling Protocol*

All cells followed the same cycling protocol and were cycled using Neware BTS4000-5V10mA (dual range) and BTS4000-5V50mA (dual range) systems. In the first cycle, cells were initially discharged with a constant current to 60% of the cathode theoretical capacity. Following this, cells were cycled at 50% of the cathode capacity with a voltage cutoff of 1.35 V when discharging, and a 2V constant voltage limit during charging to limit OER and water splitting. The *in-situ* cells were initially discharged during the first cycle at C/10 (1.25 mA/$cm^2$) and then cycled at C/5 (2.5 mA/$cm^2$) for the remainder of the cell life. The 4 $cm^2$ form factor cells were initially discharged at C/10 and then cycled at C/5 for 10 cycles before cycling at the specified rate (*i.e.* C/3 or 4.17 mA/$cm^2$). The self-discharge rate was determined by letting the 4 $cm^2$ cells rest for 1, 2, 3, and 4 weeks at room temperature (**Figure S6c, d**) before they were discharged at 2 mA (0.5 mA/cm$^2$ or C/50) to a voltage cutoff of 1 V.

The open circuit potential (OCP) measurements presented in **Figure 5** were performed using a Landt battery test system (Landt Instruments CT2001A). The cell open circuit voltage was measured every 1 hour. The DCIR used to probe the DC resistance was performed using a 5 mA/10mA pulse. The elevated temperature OCP was performed using a temperature chamber with precise control from 0 °C to 65 °C. Galvanostatic-EIS (GEIS) measurements were performed on both the *in-situ* and 4 cm$^2$ cells using a Biologic-SP150 Potentiostat. Scans were performed from 1 MHz to 1 Hz with 10 points per decade, an average of 8 measures per frequency, and an AC amplitude of 30 µA and 200 µA for *in-situ* CT cells and 4 cm$^2$ cells respectively.

Cyclic voltammetry (CV) and chronoamperometry (CA) measurements were performed in an aqueous electrolyte (10M KOH, 0.6M LiOH) in a three-electrode configuration with a Hg/HgO reference electrode and a platinum foil counter electrode. The electrochemical stability of the foils was tested first with a 1minute OCP to ensure stability of the reference electrode and was followed by 5 cycles of CV (**Figure S8**). The anode was scanned from -1.0 V to -1.5 V vs. Hg/HgO at a rate of 100 mV/s and the cathode from 0.0 V to 0.7 V vs. Hg/HgO at a rate of 10 mV/s. Following this, CA was applied for 2 minutes at the following potentials with a 1 min OCP performed in between: (1) -1.35 V, -1.39 V, and -1.45 V vs. Hg/HgO for the anode current collectors, (2) 0.2 V, 0.5 V, and 0.6 V vs. Hg/HgO for the cathode current collectors.

*X-ray CT Experimentation and Analysis*

The scans were performed with a *ZEISS Xradia 510 Versa* MicroCT instrument, and parameters were kept consistent between each scan at the various SOCs. An X-ray energy of 140 kV was used at 10 W (71.56 µA) with a voxel size of 2.504 µm, a FOV of 2.48 mm x 2.52 mm, a 1.5 hr scan time, a 4X magnification, an exposure time of 2 s, a binning of 2, and 1600 projections. The source to detector distance was also kept the same at 30.80 mm to ensure similar grey values between scans. No beam hardening correction was necessary during reconstruction. The two *in*-situ CT cells in **Figure 3** were scanned at the end of the discharge at different points of cycling: (1) after cycles 11 and 30 for the Sn-Au/SS cell, and (2) after cycles 11, 30, 51, and 101 for the Brass-Au/Ni cell. The initial or pristine case was performed on the bare anode from the same batch used in the respective cells (**Figure S2a**).

MicroCT data analysis was performed with Amira 2019 commercial software. Tomograms from the same device, but scanned at different cycles, were aligned using the "Register Images" module which uses an iterative optimization algorithm to align datasets.[53–56] After alignment, the datasets were cropped with a 600 x 600 pixel window (1.5 mm x 1.5 mm) in the XY direction to remove edge effects (**Figure S9**). With thresholding and area selection interpolation, the anode was analyzed separately by segmenting out the electrolyte and current collector regions above and below the anode in the Z direction. The volume of each anode was then analyzed, and the average thickness determined by dividing the total anode volume by the XY window.

Once segmented, the following Amira filters were applied to the reference and every anode was analyzed to ensure grey-scale values were comparable: (1) Non-Local Means (2) Unsharp Mask (3) Delineate. As shown in **Figure S3b**, once filtered, 3 distinct peaks pertaining to the Zn, ZnO, and Electrolyte-Binder-Pore (EBP) regions were visible. Global thresholding was then used to separate the three regions, and the "Volume Fraction" module was implemented to quantify the percentage of each species within the anode. The "Separate Objects" module, which implements a combination of watershed, distance transforms, and numerical reconstruction algorithms, was used on the Zn region to separate the particles.

## 3. Results and Discussion

*3.1 Current Collector Parasitic Gassing*

In the Zn-AgO system, the redox reaction relies on the dissolution of zinc and silver species in the alkaline electrolyte and their supersaturation-induced precipitation, which takes place rapidly while maintaining a stable voltage (**Equation 6-12**):[37,38,57,58]

Anode

(Dissolution) $\quad\quad\quad Zn(s) + 4OH^-(aq) \leftrightarrow Zn(OH)_4^{2-}(aq) + 2e^- \quad\quad\quad$ (6)

(Relaxation) $\quad\quad\quad Zn(OH)_4^{2-}(aq) \leftrightarrow ZnO(s) + H_2O(l) + 2OH^-(aq) \quad\quad\quad$ (7)

(Overall) $\quad\quad\quad Zn(s) + 2OH^-(aq) \leftrightarrow ZnO(s) + H_2O(l) + 2e^- \quad -1.260\ V_{SHE} \quad$ (8)

Cathode

(Dissolution) $\quad\quad\quad Ag(s) + OH^-(aq) \leftrightarrow AgOH(aq) + e^- \quad\quad\quad$ (9)

(Relaxation) $\quad\quad\quad 2AgOH(aq) \leftrightarrow Ag_2O(s) + H_2O(l) \quad\quad\quad$ (10)

(1st Transition) $\quad Ag_2O(s) + H_2O(l) + 2e^- \leftrightarrow 2Ag(s) + 2(OH)^-(aq) \quad 0.345\ V_{SHE} \quad$ (11)

(2nd Transition) $\quad 2AgO(s) + H_2O(l) + 2e^- \leftrightarrow Ag_2O(s) + 2(OH)^-(aq) \quad 0.604\ V_{SHE} \quad$ (12)

As shown in **Figure 1c,f**, the first plateau corresponds to the first phase transition from AgO to Ag$_2$O (**equation 11**), while the second corresponds to the transition from Ag$_2$O to Ag (**equation 12**). To test the effects of OER and HER of current collect materials typically used in aqZBs,[42] the current due to gassing was monitored through chronoamperometry (CA) and by applying potentials relevant to the reactions indicated above. The current density of the OER, after applying cathode-relevant potentials for 2 minutes on metal current collector rods, is presented in **Figure 2a**. Here, 0.2 V and 0.5 V vs. Hg/HgO corresponds to the first and second plateau of the cathode phase transition at 1.56 V and 1.86 V vs. Zn/ZnO respectively (**equations 11 and 12**) and can be seen in **Figure S7a** at the minima points of the Tafel plot of the cathode. Likewise, **Figure 2b** shows the HER current density after a potential hold at the anode potentials indicated in the Tafel plot of the anode in **Figure S7b**. The same method was also used to test standard metal foils, and the results are shown along with the cyclic voltammetry (CV) in **Figure S2**.

These results reveal higher OER gassing currents for bare materials than when electroplated with Au, which can also be seen in **Figure S8g** for metal foils that were sputtered with Au. However, the Au coated Ni (Au/Ni) yielded lower OER gassing currents than the Au coated Ti (Au/Ti) and Au coated stainless steel (Au/SS). On the anode end, brass and Sn rods were tested and compared. Brass rods were selected and used in the CT cells since they yielded lower HER gassing currents than copper when metal foils were tested (**Figure S8h**). Brass rods showed lower HER currents than Sn but were similar in magnitude in the higher potential region (-1.45V vs. Hg/HgO).

To test the cycle reversibility of the anode with the different current collectors, Zn symmetric CT cells with Sn and brass rods were cycled at various DoD$_{Zn}$ (**Figure 2c-d**). The anode films were initially discharged to half the desired DoD$_{Zn}$ with a potentiostat in the basic electrolyte and then assembled in symmetric cells (**Figure S4**). The first discharge was set to half the desired DoD$_{Zn}$ to fully discharge one side to the set value and charge the other back to 0%. The cell was then cycled at the full desired DoD$_{Zn}$. As expected, a general trend was observed for both cases, in which higher DoD$_{Zn}$ yielded less cyclability. The general failure mode for these cells appears to be capacity fade due to an over oxidation of Zn and a build up a ZnO. However, cell shorting was observed for the brass-brass 40% DoD shown in **Figure 2d** and **Figure S4e**. Yet, in both cases, the symmetric cells were able to cycle >100 cycles for a DoD of 10%, with the brass-brass symmetric cell achieving over 550 cycles. These results indicate good reversibility of the anode and suggest that the likely causes for failure in full cells originate at the cathode.

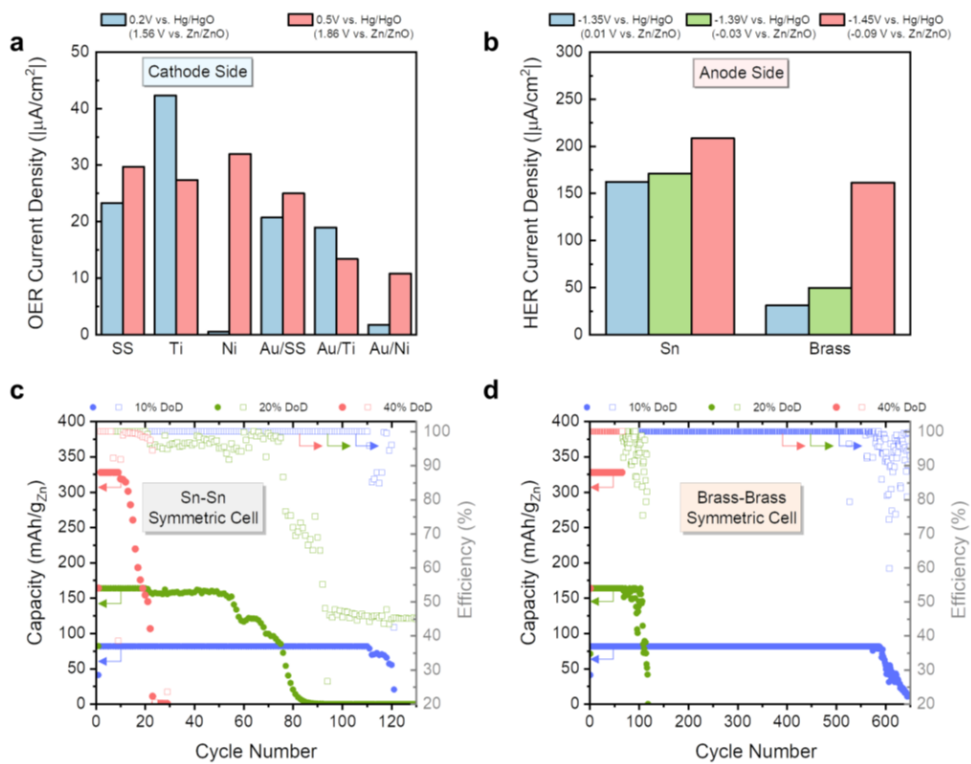

**Figure 2 | Current collector gas evolution.** Gassing currents after 2 minutes of chronoamperometry for OER at cathode relevant potentials (a) and HER at anode relevant potentials (b). Anode symmetric cell cycling performances with Sn current collectors (c) and brass current collectors (d).

*3.1 In-Situ micro-CT Analysis*

The performance of full cells was evaluated using *in-situ* MicroCT with two sets of current collectors, Sn with Au/SS rods (Sn-Au/SS) and brass with Au/Ni rods (Brass-Au/Ni), in order to determine the effects of current collector parasitic gassing. The architecture (**Figure 1d**) of the 2 mm diameter *in-situ* cells was optimized to limit common issues, such as dendrite growth (**Figure S1c-d**), so as to be representative of larger form factor performances. As shown in **Figure 3**, CT full cells were scanned *in-situ* at various points of cycling. The various phases in the anode were separated using a reference containing Zn and ZnO powder (**Figure S3**), and grey-value global thresholding was used to segment the Zn particles from the ZnO and the electrolyte-binder-pore (EBP) region as detailed in the methods section. A visualization of the segmentation process for the Sn-Au/SS case can be seen in **Video S1**.

The Sn-Au/SS CT full cell was only able to achieve ~25 cycles, while the cell with less corrosive Brass-Au/Ni current collectors lasted ~125 cycles before starting to exhibit capacity fade. The various species (Zn, ZnO, and EPB) were colorized in the XZ cross-sectional slices and 3D tomograms of **Figure 3c-d**. The quantitative analysis is presented in **Figure 4a**: it reveals an increase of ZnO with cycling for both cells, confirmed by the increasing impedance of both cells during cycling (**Figure S1f**) resulting from the buildup of the insulating ZnO. The XZ slices show a clear increase of ZnO formation (along with a decrease of Zn) near the current collectors, whereas Zn particles near the electrolyte interface are retained at longer cycle lifetimes. This disproportionate depletion of Zn and growth of ZnO near the metal interface could suggest the effects of current collector gassing or electron transport limitations in the depletion of the active material at the anode.

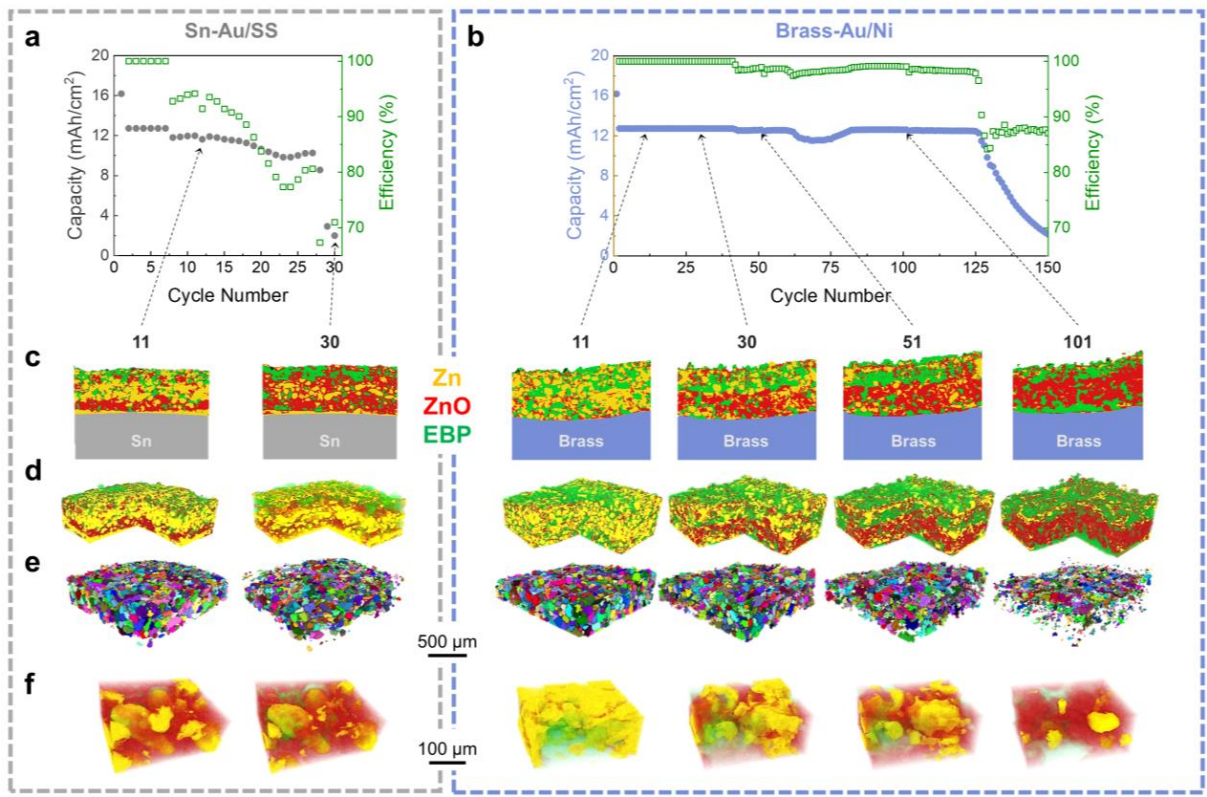

**Figure 3 | *In-Situ* MicroCT analysis.** Electrochemical cycling performance of *in-situ* MicroCT cells with (a) Sn and Au coated stainless-steel current collectors and (b) Brass and Au coated Ni current collectors. MicroCT analysis at specified cycles: (c) colorized XZ slices of Zn anode, (d) colorized corner-cut 3D reconstructed volumes of Zn anode, (e) segmented Zn particles, and (f) colorized particle scale 3D renderings. Segmentation of Zn (yellow), ZnO (red), and electrolyte-binder-pore (EBP) (green) in XZ slices and volume rendering is based on global thresholding. Zn particles in (e) are indistinctly colorized to showcase separation of individual particles.

When comparing the two current collector cases, more Zn and less ZnO is observed for the Brass-Au/Ni cell at the same cycles (*e.g.,* at cycles 11 and 30). Furthermore, the EBP volume fraction stays constant after the 11$^{th}$ cycle, showing the increase of ZnO as the main culprit for the volume expansion and the thickness increase shown in **Figure 4b**. Our results show the Zn anode volume expansion to be around 100%, which is similar to the 60-70% reported by Tobias *et al.* for an aqueous alkaline Zn-Air system.[44]

The size of the Zn particles, as depicted in **Figure 3e**, decreases with cycling. For the Sn-Au/SS, there was still a considerable amount of Zn still left after cell failure at 30 cycles. However, for the Brass-Au/Ni cell, more of the Zn was utilized, and by cycle 101, nearly all of it is converted to ZnO. This is shown in **Figure 4c**, where the equivalent diameter of the Zn particles shrinks as a function of cycling, and the Brass-Au/Ni cell can cycle longer to utilize more of the Zn particles. Fits for the particle size distribution for the two current collector cases can be seen in **Figure S10**. For the Brass-Au/Ni cell, the particles are still densely packed at cycle 11, with little ZnO, as they were when first fabricated (**Figure S2a,d**). The particle-scale tomograms (**Figure 3f**) reveal that as the cell continues to cycle, more ZnO grows around the Zn particles as the core is depleted, thus following a core-shell model as observed by Tobias et al.[44]

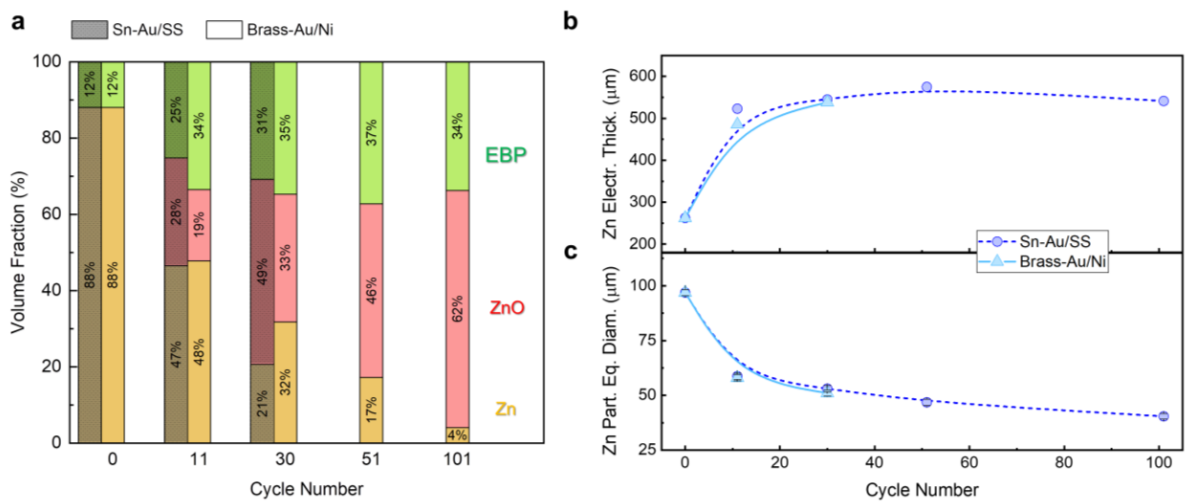

**Figure 4 | Extracted X-ray CT Statistics.** (a) Anode volume fraction of Zn (yellow), ZnO (red), and Electrolyte-Binder-Porosity EBP (green), (b) Zn anode electrode thickness, and (c) Zn particle equivalent diameter evolution during cycling. *In-situ* CT cells with Sn and Au-coated stainless-steel current collectors are represented by solid lines and triangle markers while cells Brass and Au-coated Ni current collectors are represented by dashed lines and circle markers.

*3.2 Effects of Current Collector on Shelf and Cycle Life*

The symmetric cycling of **Figure 2** suggests that the cell performance is limited by effects on the cathode side. As such, the effects of the cathode current collector on the shelf life and cycling performances were investigated for cells with Au/Ni and Au/SS current collectors. The anode current collectors were fixed to Cu, which was chosen over brass to avoid any side reaction with the electrolyte as brass contains Zn. The shelf life performances of the cells were evaluated by monitoring OCP with time at elevated temperatures. CT cells constructed with Au/Ni and Au/SS current collector rods were monitored for two weeks at 40ºC. As shown in **Figure 5a**, the Au/Ni cell exhibited less self-discharge before transitioning to the second plateau. After two weeks, CT scans were taken of the two cells, and the 3D reconstructed tomograms of the anodes were compared. The colorized XZ slices, 3D tomograms, and volume fraction percentage in **Figure 5b-c** reveal more ZnO (and less Zn) for the cell with the Au/SS current collector. Additionally, compared to the pristine case, the Au/SS cell exhibited larger Z-axis volume expansion than the Au/Ni cell. The increase of ZnO can be explained by Zn anode corrosion, originating from an increase of dissolved oxygen from the OER reaction of the cathode current collector.[40]

Ultimately, the results from the X-ray CT experimentation reveal parasitic current collector gassing to be a key degradation mechanism for both cycle and shelf life. As a result, larger 4 cm$^2$ form-factor cells were constructed and the effects of current collector parasitic gassing on electrochemical performance were examined. This was done to ensure the insights gained from CT and the smaller 2 mm *in-situ* cells extend to larger practical-scale battery systems. To test the self-discharge, shown in in **Figure 5d,** these larger form-factor cells were left to age at various temperatures while the OCP was monitored. At each temperature, the Au/SS cells fail quicker than the Au/Ni cells. The rapid drop in voltage shown for the Au/SS cell at 55ºC is likely caused by electrolyte depletion due to increased gassing at higher temperatures and is thus indicative of cell degradation or failure. However, due to less gassing, the Au/Ni cell at 55ºC demonstrated self-discharge with the expected two-phase transition. The DCIR resistance was also monitored during the OCP measurement in **Figure S6a**. The Au/SS cells quickly exhibited large increases in resistance whereas the Au/Ni cells increased slowly or were stable with time. The self-discharge rate for the Au/Ni cell at room temperature was quantified by discharging cells after controlled idle times to measure the percent of capacity loss with time. A discharge was performed every week over 4 weeks, and the results in **Figure S6c-d** reveal an average discharge rate of 7.06 %/wk, a considerable improvement with respect to previously reported lifetimes.[39,59]

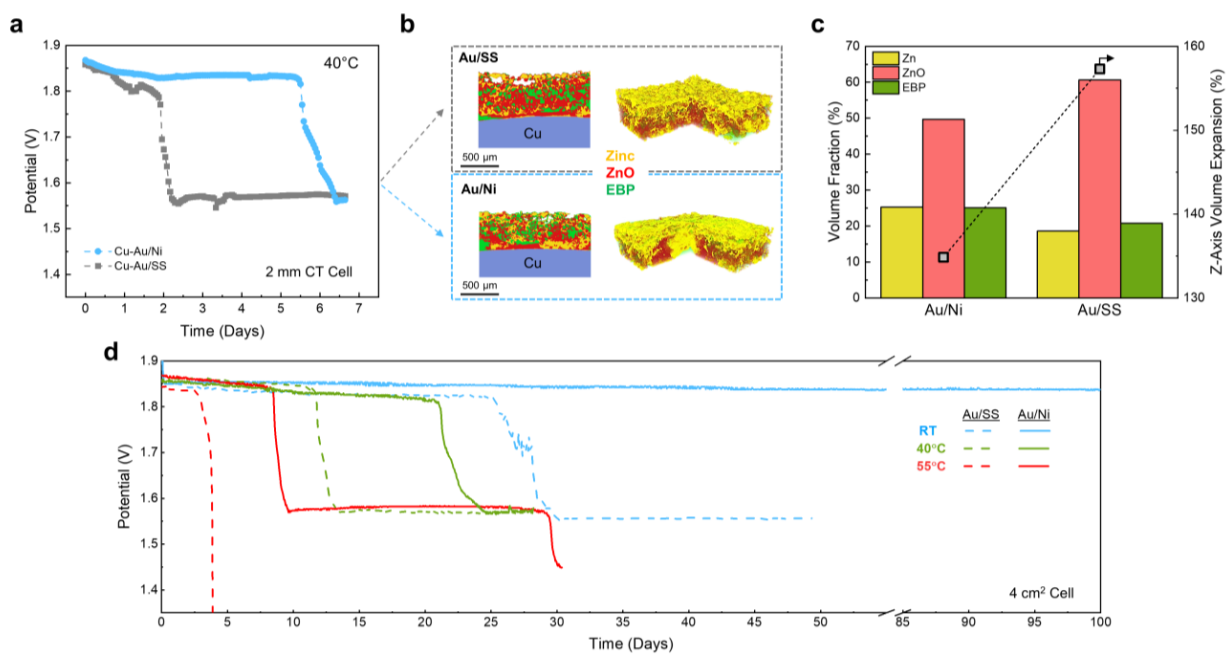

**Figure 5 | Shelf life and effect of current collector.** (a) Open-circuit voltage at 40ºC of 2 mm diameter *in-Situ* CT cell with Au coated Ni (blue) and Au coated stainless-steel (grey) cathode current collectors. (b) XZ CT slice and colorized corner-cut reconstructed volume and (c) volume fraction (*left*) of Zn (yellow), ZnO (red), and EBP (green) and z-axis volume expansion (*right*) of Zn anode after 1 week at 40ºC. (d) Open-circuit voltage of larger 4 cm$^2$ cells at room temperature, 40ºC, and 55ºC for different cathode current collectors.

The cycling performance of the cells with improved current collectors (Cu-Au/Ni) was also evaluated. **Figure 6** compares the cycling of larger form factor cells with the smaller scale *in-situ* cells fabricated for CT. The 1 cm$^2$ cell, cycling at ~8 mAh/cm$^2$ in grey, are the results of a Zn-AgO battery developed during our previous work with non-optimized current collectors.[17] Compared to this cell that demonstrated 80 cycles, both the CT and the 4 cm$^2$ cell with more electrochemically compatible current collectors demonstrated more than 250 cycles with a capacity retention exceeding 80%. The slight dip in coulombic efficiency observed, for example at around 50 cycles for the CT cell, is due to the transition between the capacity-limited regime and the voltage limited regime.[17] With the optimized current collector, the 4 cm² form factor in **Figure 6b** illustrate superior performances, >200 cycles at C/5 (2.5 mA/cm²), >325 cycles at C/3 (4.17 mA/cm²), >165 cycles at C/2 (6.25 mA/cm²). The low impedance of this cell (**Figure S6b**) with optimized current collectors allows for higher rate cycling, which is also revealed in the capacity-voltage plot in **Figure S11f** by the small IR drop when cycling at 6.25 mA/cm². Overall, the cell demonstrated a cycling volumetric energy density of 100.6 Wh/L and power density of 50.3 W/cm³.

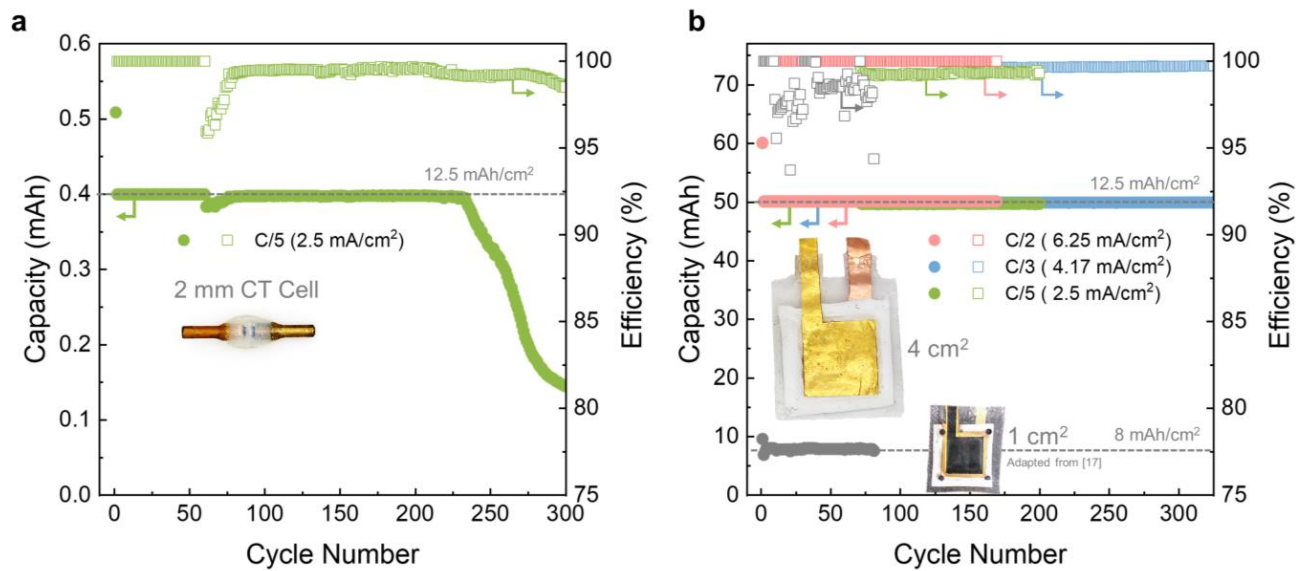

**Figure 6 | Cycle life performance.** Electrochemical cycle life of (a) 2 mm diameter *in-situ* Au/Ni-Brass CT cell and (b) larger 4 cm² Au/Ni-Cu cell and 1 cm² cell adapted from [17] with unoptimized current collectors.

## 4. Conclusion

In this work, we used *in-situ* X-ray MicroCT in combination with electrochemical experiments to investigate the effects of current collector gassing on the performances of Zn-AgO batteries. The results reveal the importance of choosing electrochemically compatible current collector materials in limiting OER and HER reactions that degrade cell performances. With *in-situ* MicroCT, we quantified the Z-axis volume expansion and showed Zn particle depletion with cycling that follows a shrinking core model. From an improved selection of electrochemically compatible current collectors, we demonstrated superior performances with a high cycling capacity of 12.5 mAh/cm$^2$ for more than 325 cycles in a 4 cm$^2$ form factor. This resulted in a cycling volumetric energy density of 100.6 Wh/L and power density of 50.3 W/cm$^3$. We have demonstrated extended cycle and shelf life enabled by the mitigation of the current collector gassing and have successfully prolonged the lifetimes of Zn/AgO batteries.

However, while we demonstrated a self-discharge rate of 7.06 %/wk, further efforts are required to extend cell shelf life further. The investigation on the Zn symmetric cells and shelf-life tests showed that the OERs at the cathode current collector are one of the limiting factors. As a result, current collector treatments and coatings need to be developed to limit the OER and improve performances. Additives in the electrodes and the electrolyte also need be explored to limit parasitic reactions, and further investigations of electrochemically compatible hydrogels could limit silver dissolution known to degrade Zn-Ag battery performances.

Lastly, through the development of representative *in-situ* CT cells, we report the first *in-situ* CT study of Zn-Ag batteries and utilize CT to study the degradation with long-term cycling (>250 cycles at 12.5 mAh/cm$^2$). The results from this work exemplify the utility and advantages of *in-situ* CT experimentation in elucidating degradation effects in battery systems. Following this work, we expect lab-scale *in-situ* CT to be increasingly used in academic and industry battery research to better improve device performances and aid in the effort to combat the energy crisis.

## Acknowledgements

This work was supported by funding from ZPower LLC, United States and Qualcomm, United States. For use of the MicroCT system, the authors would like to acknowledge the National Center for Microscopy and Imaging Research (NCMIR) technologies and instrumentation are supported by grant R24GM137200 from the National Institute of General Medical Sciences. This work was performed in part at the San Diego Nanotechnology Infrastructure (SDNI) of UCSD, NANO3, a member of the National Nanotechnology Coordinated Infrastructure, which is supported by the National Science Foundation (Grant ECCS-1542148). The authors would also like to acknowledge Neware Technology Limited for the donation of CT4000 battery cyclers, which are used to obtain the cycling data of cells in this paper.

## Conflict of Interest

The authors declare no conflict of interest.